# A Search for Early Optical Emission at Gamma-Ray Burst Locations by the Solar Mass Ejection Imager (SMEI)

Andrew Buffington[1], David L. Band[2,3], Bernard V. Jackson[1], P. Paul Hick[1], and Aaron C. Smith[1]

ABSTRACT

The Solar Mass Ejection Imager (SMEI) views nearly every point on the sky once every 102 minutes and can detect point sources as faint as R~10$^{th}$ magnitude. Therefore, SMEI can detect or provide upper limits for the optical afterglow from gamma-ray bursts in the tens of minutes after the burst when different shocked regions may emit optically. Here we provide upper limits for 58 bursts between 2003 February and 2005 April.

## 1. INTRODUCTION

Optical emission during and immediately after a gamma-ray burst may probe crucial physical processes in the evolution of these bursts, but until recently it has been poorly observed. Therefore, every additional detection and upper limit during this period is relevant to the physical understanding of the burst phenomenon. Here we report on the observations provided by the Solar Mass Ejection Imager (SMEI), a wide-angle optical telescope launched into low-Earth orbit in February, 2003. SMEI scans nearly the entire sky every orbit, viewing each sidereal location for approximately 1 minute with a 102 minute cadence. Thus, SMEI can provide a photometric measurement at the location of a gamma-ray burster within 51 minutes of the burst itself.

Gamma-ray bursts are thought to result from a relativistic jet pointed toward Earth, following the release of more than $10^{51}$ ergs in sources at cosmological distances. The flow in the jet is optically thick initially, thus concealing details of the progenitor (see Piran 2005 and references therein for details of the physical model). The gamma rays are thought to originate in "internal shocks" in this flow (Fenimore et al. 1996); except in the perhaps unusual case when the flow is uniform, regions having different Lorentz factors collide. These internal shocks may also radiate in the optical. The flow propagates out to the surrounding medium, which may be either an interstellar medium (ISM) unperturbed by the progenitor, or a circumstellar medium (CSM) produced by the progenitor, resulting in the "external shock". The X-ray-through-radio afterglow following the gamma-ray phase is attributed to synchrotron emission in the external shock (Meszaros & Rees 1997). Afterglows observed from hours to months after the burst (van Paradijs et al. 2000) are modeled by a simple relativistic blast wave. However, before the self-similar blast wave phase (the relativistic analog of the Sedov-Taylor blast wave), the interaction between the flow generating gamma-rays and the surrounding medium is characterized by a forward shock propagating into the surrounding medium and a reverse shock propagating into the burst's relativistic flow (Nakar & Piran 2004). The Lorentz factors and magnetic fields differ in these shocks, resulting in different predictions of the possible optical emission.

[1] Center for Astrophysics and Space Sciences, University of California, San Diego, CA 92093-0424; abuffington@ucsd.edu, bvjackson@ucsd.edu, pphick@ucsd.edu .
[2] GLAST SSC, Code 661, NASA/Goddard Space Flight Center, Greenbelt, MD 20771.
[3] Joint Center for Astrophysics, Physics Department, University of Maryland Baltimore County, 1000 Hilltop Circle, Baltimore, MD 21250; dband@milkyway.gsfc.nasa.gov .

Thus we may observe optical emission from:

- internal shocks during the burst;
- either the reverse or forward shock at the end of the burst; or
- the external shock propagating into the surrounding medium during the afterglow.

Therefore optical observations starting during the gamma-ray emission, and continuing until the afterglow is firmly established, probe the burst's evolution from a relativistic flow dominated by internal shocks, through the complex interaction between the flow and the surrounding medium, until a simple blast wave forms. However, obtaining the requisite optical observations during this period requires rapid detection and localization of the burst. This may be done autonomously, as with the *Swift* program (see below), or it may instead be followed by immediate notification of and quick reaction by telescopes at optical observatories. Alternatively, an optical telescope may already be observing the burst location, either serendipitously or by design, using a wide field-of-view (FOV) telescope to monitor the gamma-ray burst's sky location.

Robotic telescopes such as the Livermore Optical Transient Imaging System (LOTIS--Williams et al. 1999) and the Robotic Optical Transient Search Experiment (ROTSE--Akerlof et al. 2000) provided some of the first detections and upper limits of the optical emission in the minutes after a burst. The first detection of optical emission during a burst was by ROTSE during the burst GRB 990123 (Akerlof et al. 1999). Here, the optical lightcurve increased to V=8.9 and then decreased, and was uncorrelated with the gamma rays (Briggs et al. 1999). Kulkarni et al. (1999) also detected a radio flare a day after this burst, supporting the hypothesis that the optical emission was due to synchrotron emission from the reverse shock.

More recently, optical emission (Vestrand et al. 2005) during, and infrared (Blake et al. 2005) during and immediately after, the burst GRB 041219A was detected. One of the Rapid Telescopes for Optical Response (RAPTOR) began observing the burst a little over 100 s after the burst trigger, which was a precursor 250 s before the burst's main peak. Therefore, RAPTOR followed the optical emission during the burst's brightest emission; the optical emission peaked at an extinction-corrected Rc~13.7. Although the RAPTOR observations lacked time resolution, they are consistent with being correlated with the gamma-ray emission, contrary to the optical-gamma-ray relationship during GRB 990123 (Vestrand et al. 2005). The maximum infrared fluxes (J~14, H~15 and Ks~16) were observed when the Peters Automated Infrared Imaging Telescope (PAIRITEL) slewed to the burst location 400 s after the burst began, while the burst was still in progress. During the following 3000 s the infrared flux varied, suggesting the lightcurve peaks three times, perhaps resulting from some combination of internal shocks and the forward and reverse shocks from the external shock region (Blake et al. 2005). Based on both the RAPTOR and PAIRITEL observations, Vestrand et al. (2005) suggest that: the gamma-ray spectrum extends to optical wavelengths during the burst; the reverse and forward shocks during the afterglow's early phase results produce brighter optical-infrared emission; and the late afterglow results in fading optical emission.

Launched in November 2004, the *Swift* observatory (Gehrels et al. 2004) detects and localizes bursts with its Burst Alert Telescope (BAT), a 15-150 keV coded-mask detector, and then slews to place the burst within the FOV of the onboard X-ray Telescope (XRT) and UV-Optical Telescope (UVOT) which are co-aligned with the center of the BAT's much larger FOV. Thus the UVOT on *Swift* should observe early optical emission from a large number of gamma-ray bursts.

However, because of operational constraints the UVOT will not always be able to observe each burst *Swift* detects within the first few minutes of the burst, nor will *Swift* detect every burst in the sky.

SMEI (see next section) is a spaceborne optical monitor having a FOV of $3° \times 160°$ which scans over nearly the entire sky every orbit. SMEI's primary objective is the observation of mass ejections from the Sun by detecting large-spatial-scale increases in the optical surface brightness caused by Thomson scattering of sunlight upon free electrons in the interplanetary medium. This large-scale surface brightness must be measured against the much brighter (by typically two orders of magnitude) stellar and zodiacal-light optical field. The design photometric precision required for these solar observations allows SMEI to detect (at roughly $3\sigma$) new point sources equivalent to $V=10^{th}$ magnitude.

This paper reports upper limits for optical emission at the locations of 58 out of 91 gamma-ray bursts detected since SMEI began normal operations in early February 2003, through a cutoff for the present work of 2005 April 6. We employ a one-orbit time window extending from 51 minutes before each burst until 51 minutes after. The particular full-sky map viewing the gamma-ray burst location at the time within this window thus includes the photometric measurement of that location, at the closest time available from SMEI. We then compare this measurement with similar ones from one orbit earlier and one orbit later.

These $10^{th}$ magnitude upper limits from SMEI are not particularly deep, but are nonetheless relevant. The optical emission detected from GRB 990123 peaked at V=8.9, although other bursts had much fainter upper limits, both in the observed magnitude and when scaled to the burst intensity. For most of the time period covered in this paper, there were few operating optical telescopes that observed burst locations within minutes. Even with *Swift* now being operational, SMEI may sometimes provide an earliest optical observation because the UVOT may not be able to reach the burst location immediately; *Swift* may not be able to relay the information to the ground (e.g., the TDRSS link may be disrupted); and ground-based optical telescopes may not be able to observe because of proximity to the Sun or poor weather. In addition, when the timing is fortunate, SMEI may cover the gamma-ray burst sky location in the minutes before the observation of the burst. Thus SMEI can continue to provide relevant observations.

The next two sections describe SMEI, the nature of its data, and its analysis. The section following these presents results, and the final section a conclusion and discussion.

## 2. THE SOLAR MASS EJECTION IMAGER

The Solar Mass Ejection Imager (SMEI) is designed to detect and forecast the arrival at Earth of solar mass ejections and other heliospheric structures present in the inner heliosphere. SMEI is mounted on the *CORIOLIS* spacecraft as shown in figure 1. Each camera views the sky through a stray-light reducing baffle and subsequent all-reflecting optics. For more details about the instrument see Eyles et al. (2003), and about the SMEI mission in Jackson et al. (2004). The aperture area is 1.76 $cm^2$ and the bandpass that of a wide-open CCD detector, thus corresponding roughly to the "R" photometric band, but having a twice wider bandwidth. A $10^{th}$ magnitude G-type star viewed on-axis produces approximately 800 photoelectrons during a single 4-second exposure. A sidereal location near the orbital plane (roughly over the Earth's terminator) remains within the unvignetted FOV for about 1 minute each 102 minutes, while regions closer to the orbit's poles (towards the solar and anti-solar directions) are observed for longer.

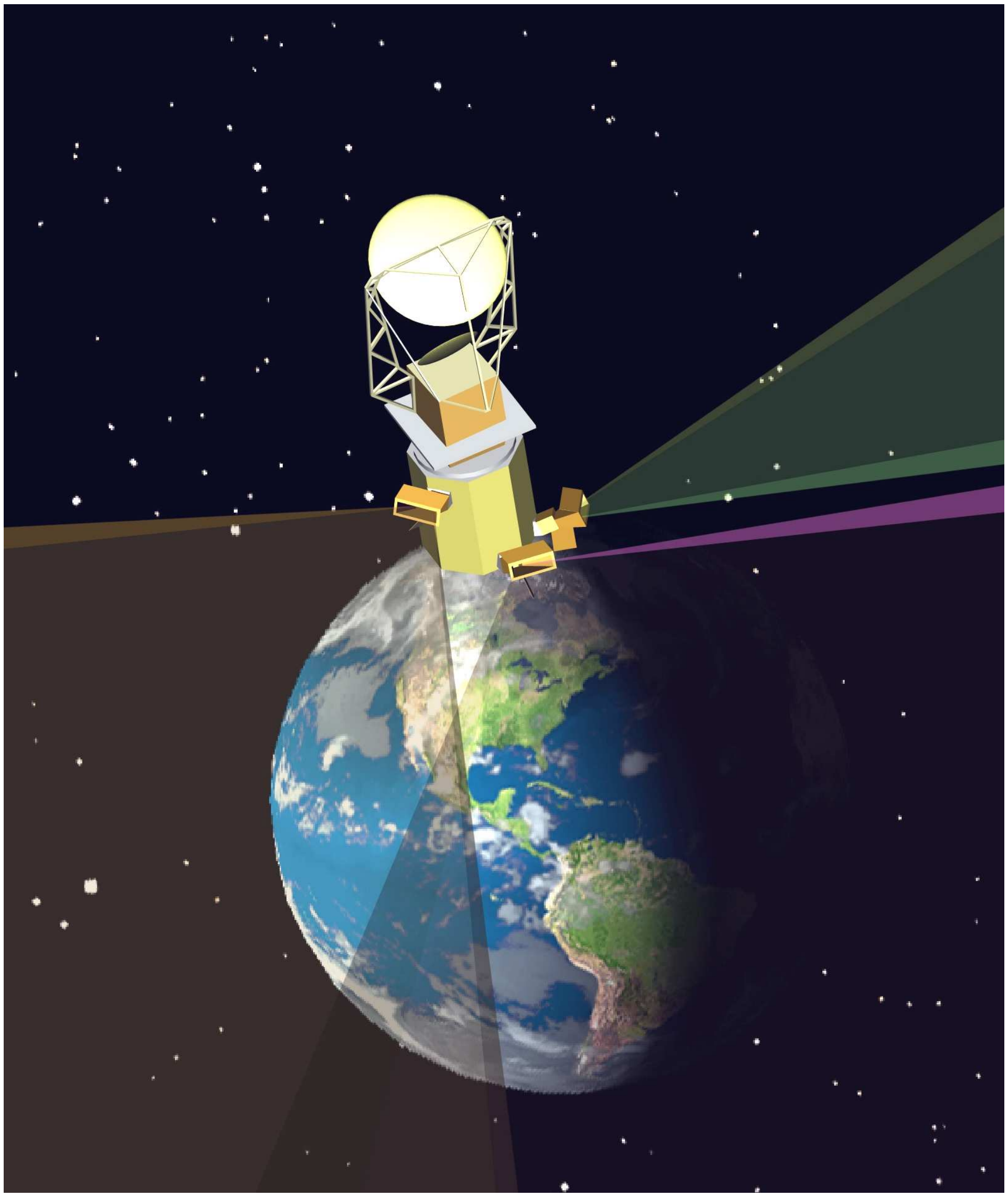

Fig. 1. – Schematic of SMEI as deployed in orbit on the *CORIOLIS* spacecraft, having just reached its northernmost position, after traveling north above the dusk terminator. The orbit is nearly circular at 840 km above the surface of the Earth, and has an inclination of 98° relative to the equatorial plane. The spacecraft is zenith pointing and oriented as shown with always the same side towards the Sun. SMEI looks outward over approximately a 160° range of sky using three strongly-baffled cameras. Their fields of view, depicted above as colored viewing columns, are oriented approximately 30° above the local horizontal to avoid both light from the Earth and sunlight from the rotating *Windsat* antenna, the other instrument on board this satellite. Each camera has a 3° × 60° FOV: together these sweep out nearly the entire sky beyond about 18° from the Sun. In this article, the SMEI cameras are numbered from 1, viewing farthest from the Sun, to 3, viewing closest to the Sun.

## 3. SMEI DATA AND ITS ANALYSIS

Figure 2 shows a typical SMEI data frame, a single 4-second exposure starting at 13:04:47 UT on 2004 March 23 from camera 2. A gamma-ray burst was observed by *INTEGRAL* (Mereghetti et al. 2004) at the indicated location just one minute earlier. This event has the closest match between burst time and SMEI observation, of the burst locations studied in this paper. The brightness unit here is an analog-to-digital unit (ADU), one of which on the CCD is 4.7 electrons detected by a single pixel, which in turn on average covers a 0.05° × 0.05° patch on the sky (Eyles et al. 2003). This average brightness scale is retained when the CCD pixels are 4×4 (cameras 1 & 2) or 2×2 (camera 3) binned for telemetry onboard the satellite. The UCSD SMEI data analysis further retains this average surface brightness when combining a sequence of frames to make a sky map (Jackson et al. 2004). The data analysis also removes most (but not all) of the Earth-radiation-band individual-pixel particle hits on the CCD, some of which are visible in figure 2. Figure 3 presents a sky map using a sequence that spans a generous time window around the individual frame of figure 2. Calibration analysis is presently underway at UCSD, using hundreds of stars brighter than about 4.5th magnitude, to establish a definitive relationship between the ADU scale on maps such as figure 3, and an “S10,” which is the equivalent surface brightness of a 10th magnitude G-type star, spread over a 1° × 1° sky bin. The present value is 1 S10 = 0.4 ADUs, somewhat smaller than the preliminary 0.55 value of figure 8 in Jackson et al. (2004).

SMEI suffered a data-handling-unit reset about a half hour after the data sequence shown in figure 3, which was not remedied until about 10 hours later, so this particular gamma-ray-burst sky location is covered only up to the orbit shown in figure 3. As expected, the one-orbit-earlier equivalent map to figure 3 is nearly identical. Figure 4 is the difference map between these two, and it shows no visible-light evidence of a burst at this location. The next section summarizes this result quantitatively, and presents the results of similar analyses for the other gamma-ray-burst events covered in the present work.

We considered a list of 91 classical bursts (i.e., we excluded Soft Gamma-ray Repeaters) that were detected between 2003 February 6 when SMEI began regular operations and 2005 April 6. This list was selected from J. Greiner’s burst compilation[4], with a few additional bursts from the *HETE II* [5] and *Swift* [6] burst databases; data describing these bursts were extracted from all three databases. Of these, SMEI did not take data during 9, was in calibration mode for 22, and had data ruined by South-Atlantic-Anomaly particle CCD hits for two. This leaves 58 surviving candidates whose sky locations were covered within ± ½ orbit (51 minutes) by SMEI. Table 1 lists these bursts along with the burst time, location and fluence, and the detecting mission. Fluences are not available for the bursts observed by *INTEGRAL* and the *IPN,* as well as for some faint bursts detected by *HETE II* and *Swift*.

Orbit-to-orbit difference maps like figure 4 are formed for each candidate. The selection of SMEI data frames is chosen generously (*i.e.*, for about triple the time duration that the GRB location remains within the FOV) to cover the region of sky containing the gamma-ray burst, for the orbit nearest in time to the burst, and the orbits immediately preceding and following it. Data are then combined within 1° × 1° sky bins, one centered upon the expected gamma-ray burst location and its eight neighboring bins. This sky bin size covers the fish-shaped triangular point spread function for SMEI (see section 3.1 in Eyles et al. [2003] for a more detailed description of the optics). Figure 5 shows the appropriate portion of figure 4 enlarged and converted to RA and δ co-ordinates, and the appropriate nine bins above shown as boxes. A plane is fitted through the values of the eight neighboring bins, and then the values predicted by this plane are subtracted from the observed values in all nine bins. This process removes any potential transient difference between the maps, such as auroral light or heliospheric structures, which varies slowly with sky position.

[4] http://www.mpe.mpg.de/~jcg/grbgen.html
[5] http://space.mit.edu/HETE/Bursts
[6] http://swift.gsfc.nasa.gov

Next, a standard deviation $\sigma = \sqrt{\sigma^2}$ is calculated using the neighbors alone, from the following variance formula:

$$\sigma^2 = \frac{1}{5} \sum_{i=1}^{8} \left( R_i - R_i^p \right)^2 . \tag{1}$$

Here $R_i$ and $R_i^p$ are respectively the differential photometric response for the eight bins, the fitted planar value and the 5 in the denominator is the number of degrees of freedom (8 measurements minus the 3 parameters for the plane). This $\sigma$ estimates the intrinsic noise at this sky location, under the assumption that the intrinsic noise of a given 1° × 1° sky bin differs little from its immediate neighbors. This assumption is likely correct for additional noise such as slowly varying background light or a residue of particle hits not removed by the analysis. Some of the latter are seen in figure 4 as isolated red and blue spots, and in figure 5 as patchy red and blue regions smaller than the stars. On the other hand, this assumption may not be correct when a nearby bright star lies partly or wholly within the group of nine bins.

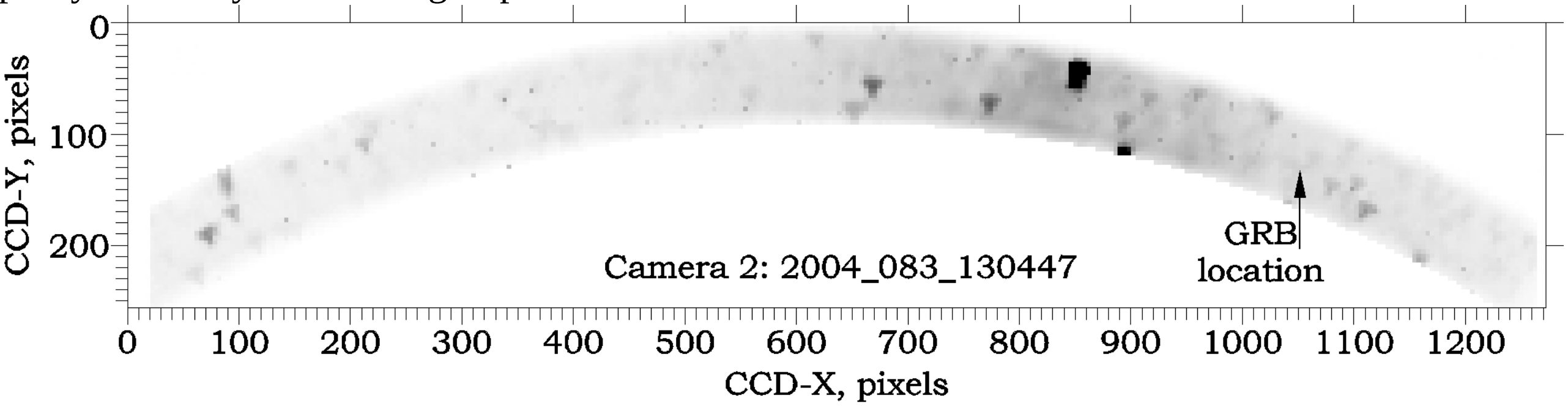

Fig. 2. –A data frame from SMEI camera 2 recorded on 2004 March 23, at 13:04:47 UT. The CCD pixel size labeling the axes here is about 0.05° × 0.05° on the sky, but these pixels were 4 × 4 binned onboard the satellite to reduce the amount of telemetry. Contrast has been enhanced so white-to-black here represents a span of 1000 ADUs instead of the full dynamic range of ~ 65,000. An electronic pedestal offset and dark current have been removed. The time is 108 s after *INTEGRAL*'s observation of GRB 040323: in this time SMEI moved to center the gamma-ray burst sky location in the FOV narrow dimension. In a sequence of frames, stars travel across the FOV view approximately radially along the narrow dimension. Sky viewed here extends from [R.A., δ] = [$21^h30^m$, -65°] at the left of this frame, to [$13^h40^m$, -43°] at the right. The optical design bends a straight line on the sky into the above arc. The star α Cen is here located in the middle of the dark Milky Way band, just to center right above; having a peak intensity of 12,000 ADUs, this star is greatly saturated with this contrast enhancement. The arrow marks the expected GRB location. Just to the right of this is the triangular group formed by M Cen (closest, about $5^{th}$ magnitude), V774, and (furthest) a combination of 4 stars including SAO241177. SMEI's triangular-shaped point spread function is evident here, as is a population of scattered particle hits on the CCD, from Earth radiation bands.

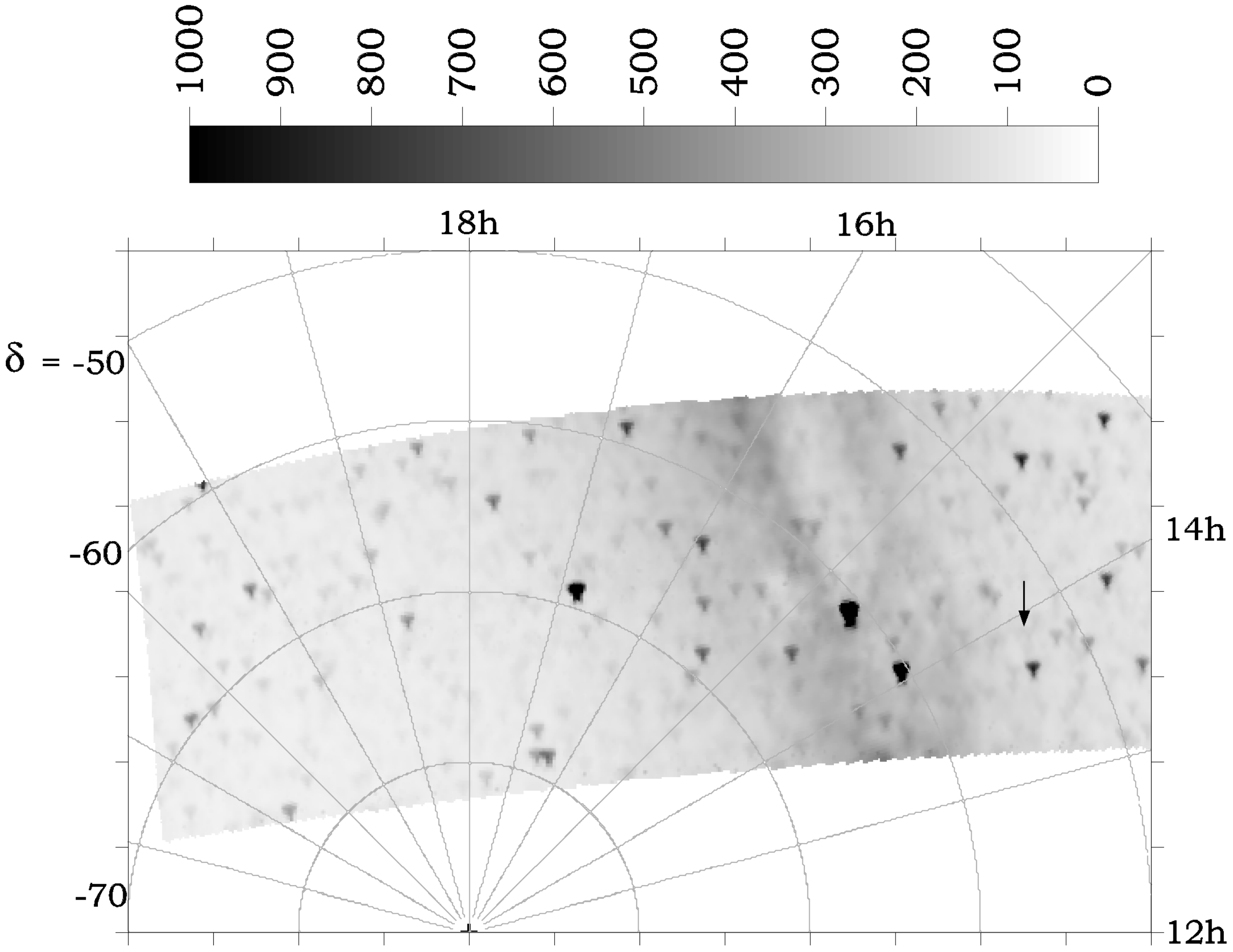

Fig. 3. –Combined average of 82 SMEI data frames from 13:01:03 to 13:06:27 UT on 2004 March 23. The scale here and in the next two figures is analog-to-digital units (ADUs) of sky surface brightness, where the original data frame brightness scale (ADUs per [0.05° × 0.05°] CCD pixel) is retained, see text. The equivalent surface brightness of a $10^{th}$ magnitude star here is about 0.4 ADUs. The coordinate frame is Right Ascension (hours, right and top axes) and Declination (degrees, left axis). The arrow marks the sky location of the gamma-ray burst.

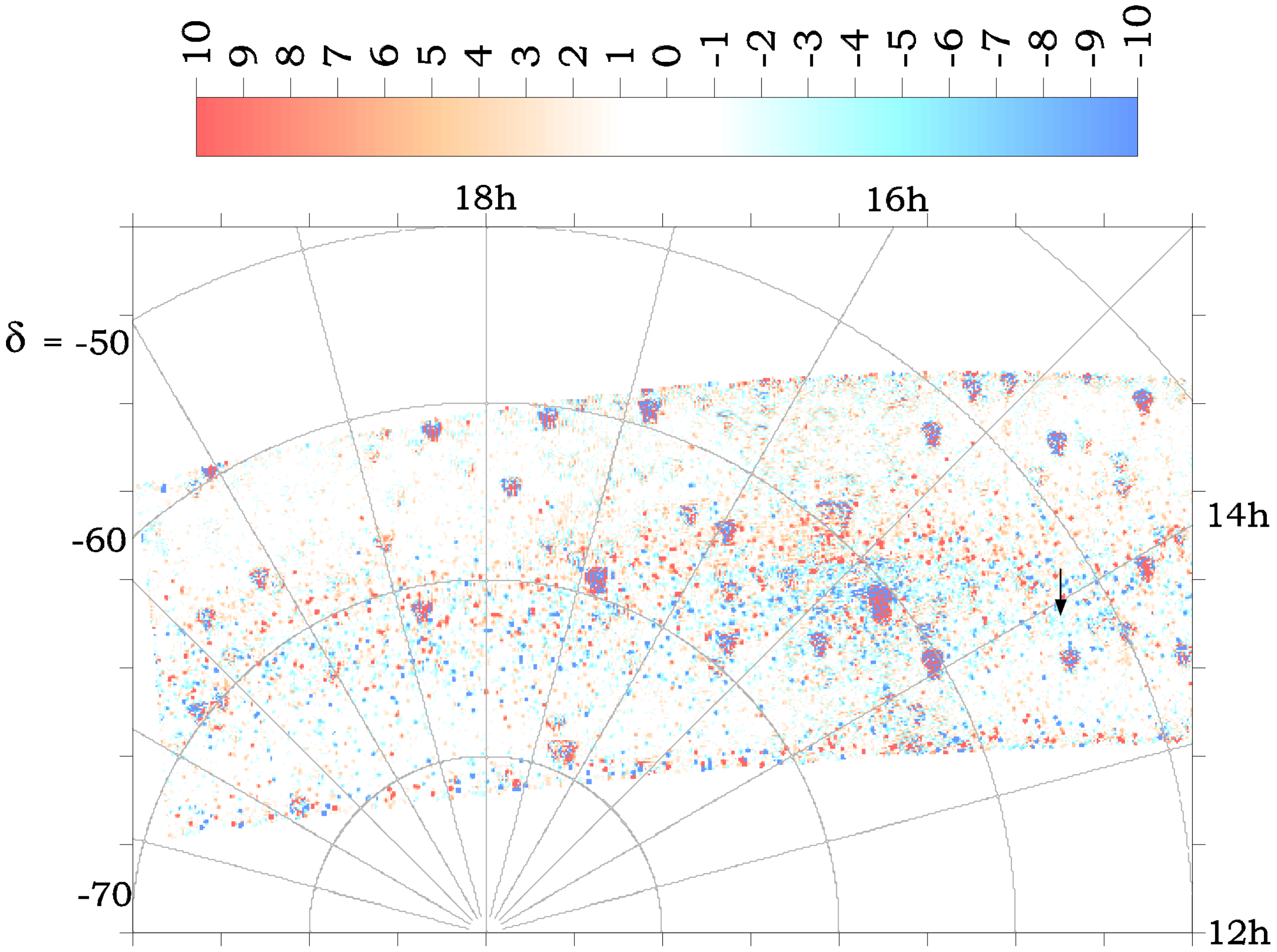

Fig. 4. –The sky map shown in figure 3 minus the equivalent map for the immediately previous orbit. The difference intensity scale (ADUs) is 100× enhanced compared with that of figure 3. As expected, the subtraction becomes noisy near bright stars. Away from these, a majority of difference bins are smaller than ±1 ADU. A residue of auroral-band particle contamination, and perhaps even some auroral light on the CCD can be seen as the red (above) and blue (below) band that runs roughly horizontally across the middle of this figure. The arrow, as in figure 3, marks the location of the gamma ray burst. The slight preponderance of blue (negative values) here indicates that no increase of light is observed at this location immediately after the burst.

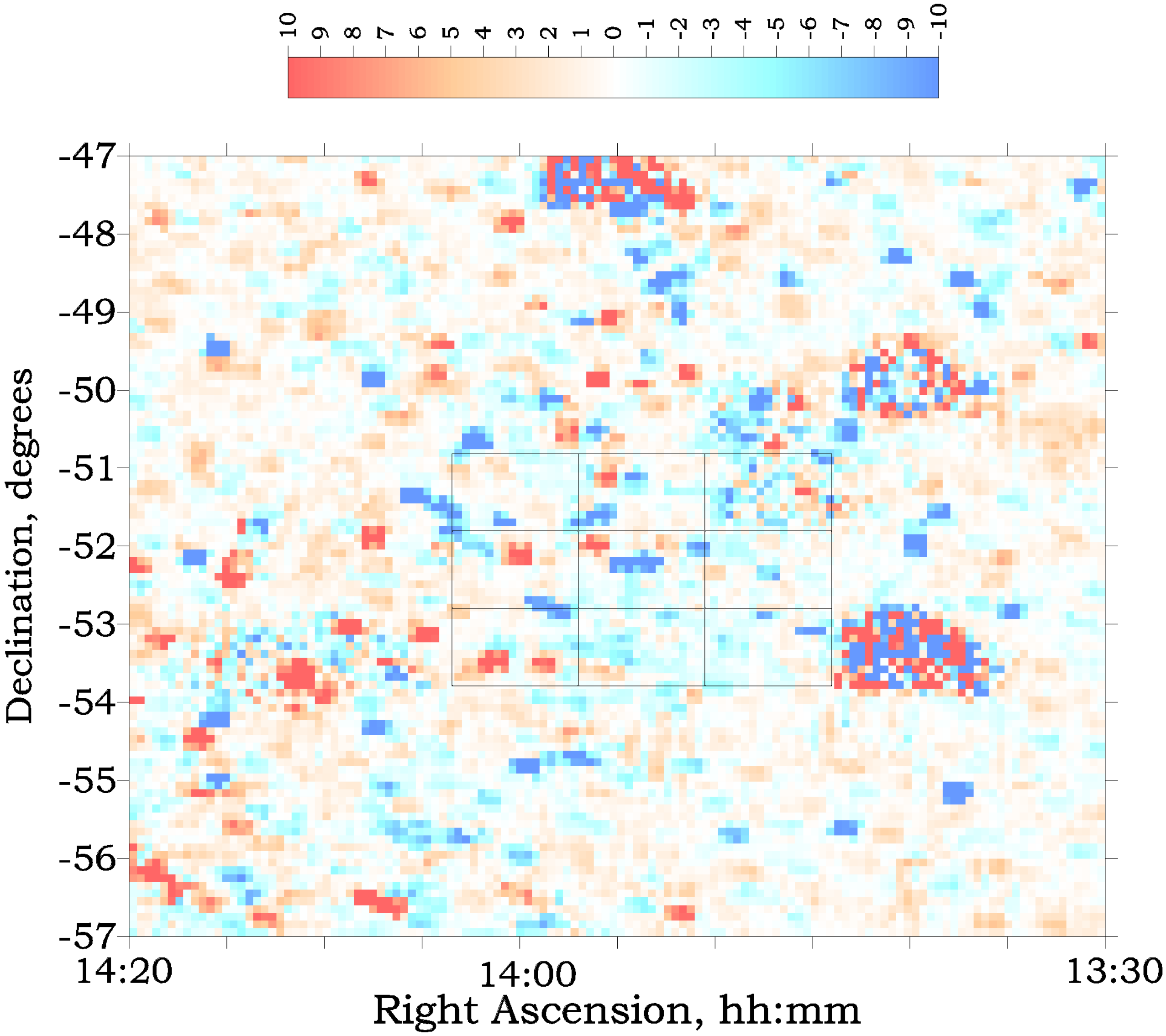

Fig. 5. –The portion of figure 4 near the gamma-ray burst enlarged and with coordinates converted to RA and δ. Also shown are the nine sky bins described in the text which are used to evaluate the presence or absence of an optical counterpart to the gamma-ray burst. At this enlargement, an optical counterpart would show as a red region occupying (and mostly filling) the center bin. Instead, we see smaller regions of red and blue, typical residues of particle impacts on the CCD, not entirely removed by the indexing analysis. At this enlargement also, the subtraction of individual stars such as bright ε Cen, located here just to the right of the boxes, is noisy because of photoelectron counting-statistics uncertainties within these small sky bins.

Five of these bursts have sky locations uncertain by more than 0.5°, see column 6 in table 1. For these a more extensive analysis was performed to see whether relocating the analysis region at 0.1° intervals, to cover a 2° × 2° square centered on the nominal expected sky location, would alter the result. Photometric results for three of these were insensitive to this change in the expected location of the GRB. Two showed a photometric change, but at the nearby location of a presumably variable bright star; these thus are discounted here as "GRBs". GRB 040425 received a further analysis, because its error box is 3' × 7° (*i.e.,* long and narrow) in the sky: here we repeated the above, but using larger 0.3° intervals to explore a 6° × 6° square of sky. The photometric result remained essentially unchanged, except at several bright-star locations contained within this larger area.

## 4. RESULTS

Table 1 lists the 58 gamma-ray bursts, and summarizes their properties. Table 2 provides the SMEI results for the same 58 bursts. Columns 6 and 7 respectively provide the differences between the closest-in-time SMEI measurement and the one-orbit-earlier and one-orbit-later SMEI measurements, and columns 8 and 9 the neighbors-defined standard deviation. Columns 10 and 11 respectively provide the ratio of columns 6 to 8 and 7 to 9 (*i.e.,* the signal-to-noise ratio, hence a measure of the statistical significance of the measurement). Briefly, no optical counterparts are observed greater than 5.3 standard deviations, for any of the bursts.

TABLE 1
Gamma-ray burst locations that were also observed by SMEI within ± ½ orbit

| # | Name | Time, UT | R.A. | Dec. δ | Δ Pos[a] ( ' ) | Mission | Fluence[b] |
|---|---|---|---|---|---|---|---|
| 1 | GRB 030320 | 10:11:40 | 17:51:31 | -25°18' | 5 | *INTEGRAL* | |
| 2 | GRB 030324 | 03:12:43 | 13:37:11 | -00°19' | 7 | *HETE* | $12.75^{+3.35}_{-3.01}$ |
| 3 | GRB 030325 | 14:15:10 | 04:43:07 | -19°06' | 60×4 | *IPN* | |
| 4 | GRB 030326 | 10:43:41 | 19:31:52 | -11°43' | 65×3 | *IPN* | |
| 5 | GRB 030328 | 11:20:58 | 12:10:46 | -09°22' | 2 | *HETE* | $287.4^{+13.9}_{-14.1}$ |
| 6 | GRB 030329 | 11:37:15 | 10:44:50 | +21°31' | 2 | *HETE* | $1076.^{+13.}_{-14.}$ |
| 7 | GRB 030418 | 09:59:19 | 10:54:53 | -06°59' | 14 | *HETE* | $17.34^{+7.27}_{-5.22}$ |
| 8 | GRB 030429 | 10:42:23 | 12:13:06 | -20°56' | 2 | *HETE* | $3.80^{+1.40}_{-1.17}$ |
| 9 | GRB 030519 | 14:04:54 | 16:04:36 | -33°28' | 17×2 | *HETE* | 609.0±9.7 |
| 10 | GRB 030528 | 13:03:03 | 17:04:02 | -22°39' | 2 | *HETE* | $56.34^{+7.13}_{-7.32}$ |
| 11 | GRB 030723 | 06:28:17 | 21:49:30 | -27°42' | 2 | *HETE* | $0.38^{+5.56}_{-0.33}$ |
| 12 | GRB 030725 | 11:46:25 | 20:33:47 | -50°46' | 28 | *HETE* | $166.7^{+10.3}_{-10.1}$ |
| 13 | GRB 030821 | 05:31:36 | 21:42:33 | -45°12' | 120×10 | *HETE* | $27.47^{+3.35}_{-2.99}$ |
| 14 | GRB 030823 | 08:52:41 | 21:30:47 | +21°59' | 6 | *HETE* | $12.74^{+4.43}_{-3.53}$ |
| 15 | GRB 030824 | 16:47:35 | 00:05:02 | +19°55' | 11 | *HETE* | $5.83^{+2.38}_{-1.89}$ |
| 16 | GRB 030913 | 17:06:58 | 20:58:02 | -02°12' | 30 | *HETE* | $8.04^{+2.69}_{-1.93}$ |
| 17 | GRB 031026 | 05:35:43 | 03:18:42 | +28°22' | 15 | *HETE* | 28 |
| 18 | GRB 031203 | 22:01:28 | 08:02:30 | -39°51' | 2.5 | *INTEGRAL* | |
| 19 | GRB 031220 | 03:29:57 | 04:39:34 | +07°22' | 11 | *HETE* | 4 |
| 20 | GRB 040106 | 17:55:12 | 11:52:15 | -46°46' | 3 | *INTEGRAL* | |
| 21 | GRB 040323 | 13:02:59 | 13:53:53 | -52°21' | 2 | *INTEGRAL* | |
| 22 | GRB 040403 | 05:08:08 | 07:40:55 | +68°13' | 2.7 | *INTEGRAL* | |
| 23 | GRB 040425 | 16:25:10 | 15:31:36 | -39°43' | 3×420 | *HETE* | 159 |
| 24 | GRB 040511 | 13:01:46 | 14:47:50 | -44°15' | 1.3 | *HETE* | 320 |
| 25 | GRB 040624 | 08:21:35 | 13:00:10 | -03°35' | 3 | *INTEGRAL* | |
| 26 | GRB 040701 | 13:00:55 | 20:47:46 | -40°14’ | 8 | *HETE* | |
| 27 | GRB 040709 | 00:58:08 | 20:53:53 | -28°13’ | 95 | *HETE* | |
| 28 | GRB 040730 | 02:11:55 | 15:53:16 | -56°27’ | 2 | *INTEGRAL* | |
| 29 | GRB 040810 | 14:15:36 | 23:54:08 | -35°04’ | 30 | *HETE* | 350 |
| 30 | GRB 040812 | 06:01:50 | 16:26:05 | -44°43’ | 2 | *INTEGRAL* | |
| 31 | GRB 040825A | 03:30:30 | 22:58:59 | -10°56’ | 7 | *HETE* | |
| 32 | GRB 040825B | 16:21:37 | 22:46:34 | -02°24’ | 9 | *HETE* | |
| 33 | GRB 040903 | 18:17:55 | 18:03:22 | -25°15’ | 2.5 | *INTEGRAL* | |
| 34 | GRB 040912 | 14:12:17 | 23:56:54 | -01°00’ | 7 | *HETE* | 23 |
| 35 | GRB 040916 | 00:03:30 | 23:01:30 | -05°35’ | 17 | *HETE* | 7.7 |
| 36 | GRB 040924 | 11:52:11 | 02:06:19 | +16°01’ | 6 | *HETE* | 94. |
| 37 | GRB 041006 | 12:18:08 | 00:54:53 | +01°12’ | 5 | *HETE* | 290. |
| 38 | GRB 041217 | 07:28:30 | 10:59:10 | -17°57’ | 12 | *Swift* | 65.7 |
| 39 | GRB 041218 | 15:45:25 | 01:39:06 | +71°20’ | 2.5 | *INTEGRAL* | |
| 40 | GRB 041219A | 01:42:18 | 00:24:26 | +62°50’ | 2 | *INTEGRAL/Swift* | 1000 |
| 41 | GRB 041219B | 15:38:48 | 11:10:42 | -33°27’ | 12 | *Swift* | |
| 42 | GRB 041224 | 20:20:57 | 03:44:48 | -06°39’ | 7 | *Swift* | 125±6 |
| 43 | GRB 041226 | 20:34:19 | 05:18:10 | +73°21’ | 6 | *Swift* | 5.87±2.71 |
| 44 | GRB 041228 | 10:49:13 | 22:26:34 | +05°03’ | 5 | *Swift* | 51.4±3.4 |
| 45 | GRB 050117A | 12:52:36 | 23:53:42 | +65°56’ | 4 | *Swift* | 156±6 |
| 46 | GRB 050117B | 20:03:35 | 16:13:31 | -70°21’ | 10 | *Swift* | |
| 47 | GRB 050123 | 10:22:53 | 10:31:34 | -11°34’ | 14 | *HETE* | 31 |
| 48 | GRB 050124 | 11:30:03 | 12:51:31 | +13°02’ | 6 | *Swift* | 20.0±1.4 |
| 49 | GRB 050128 | 04:20:04 | 14:38:21 | -34°46’ | 3 | *Swift* | 85.0±5.3 |
| 50 | GRB 050129 | 20:03:03 | 16:51:12 | -03°05’ | 3 | *INTEGRAL* | |
| 51 | GRB 050219A | 12:40:01 | 11:05:38 | -40°41’ | 4 | *Swift* | 76.9±4.0 |
| 52 | GRB 050219B | 21:05:51 | 05:25:06 | -57°46 | 4 | *Swift* | 222±10 |
| 53 | GRB 050306 | 03:33:12 | 18:49:17 | -09°10’ | 4 | *Swift* | 184±10 |
| 54 | GRB 050315 | 20:59:43 | 20:25:53 | -42°35’ | 4 | *Swift* | 33.9±2.1 |
| 55 | GRB 050318 | 15:44:37 | 03:18:43 | -46°24’ | 4 | *Swift* | 16.5±1.2 |
| 56 | GRB 050319 | 09:31:18 | 10:16:38 | +43°34’ | 6 | *Swift* | 6.37±0.92 |
| 57 | GRB 050326 | 09:53:56 | 00:27:26 | -71°22’ | 4 | *Swift* | 178±6 |
| 58 | GRB 050401 | 14:20:15 | 16:31:31 | +02°11’ | 6 | *Swift* | 134±8 |

[a] The uncertainty in the position, usually the 90% confidence radius.

[b] Burst energy fluence F in units of $10^{-7}$ erg $cm^{-2}$. The *HETE II* fluences are for the 30-400 keV band while the *Swift* fluences are for the 25-350 keV band. *HETE II* fluences until GRB 030913 are taken from Sakamoto et al. (2005). Subsequent *HETE II* bursts are from pipeline processing reported on the *HETE II* website[5], and were recalculated for the 30-400 keV band using spectral fits also reported on the website. The *Swift* fluences result from pipeline processing. No fluences are available for the bursts detected by *INTEGRAL* and the *IPN*, and for some faint *HETE II* and *Swift* bursts.

In Table 2, note when Δt is positive that SMEI looks at that particular place in the sky *after* the GRB time and thus an optical counterpart would be indicated by a positive value in the “Orbit 2-1” column. When Δt is negative, SMEI looks earlier, so a positive value in the “Orbit 2-1” would indicate that an optical counterpart had occurred *before* the indicated GRB time. If such optical counterpart were to fade quickly, this would be followed by a comparable negative value in the “Orbit 3-2” column. If on the other hand with Δt negative, an optical counterpart were to be present only in the third orbit, we would expect then to see a small or zero value in the “Orbit 2-1” column, and instead a positive value in the “Orbit 3-2” column.

TABLE 2

SMEI measurements at gamma-ray burst locations

| # | Name | Cam # | $\Delta t^a$ | $N^b$ | $\Delta_{2-1}$ (ADUs) | $\Delta_{3-2}$ (ADUs) | $\sigma_{21}$ | $\sigma_{32}$ | $\frac{\Delta_{2-1}}{\sigma_{21}}$ | $\frac{\Delta_{3-2}}{\sigma_{32}}$ | 3σ limit $M_v$ (magnitudes) | Comments[c,d,e,f] |
|---|---|---|---|---|---|---|---|---|---|---|---|---|
| 1 | GRB 030320 | 2 | +35 | 21 | 3.65 | 12.3 | 2.39 | 6.5 | 1.5 | 1.9 | 6.6 | d |
| 2 | GRB 030324 | 1 | -28 | 49 | 0.13 | 0.53 | 0.49 | 0.25 | 0.3 | 2.1 | 9.3 | c |
| 3 | GRB 030325 | 2 | +45 | 21 | -0.83 | 1.62 | 0.81 | 0.44 | -1.0 | 3.7 | 8.7 | e |
| 4 | GRB 030326 | 2 | -4 | 21 | -0.14 | 0.06 | 0.05 | 0.24 | -2.8 | 0.3 | 10.3 | |
| | | 3 | -7 | | 0.48 | 0.17 | 1.98 | 0.33 | 0.2 | 0.5 | ~9.0 | Orbit 1: Moon nearby |
| 5 | GRB 030328 | 1 | +44 | 68 | 0.26 | 0.35 | 0.70 | 0.24 | 0.4 | 1.5 | 9.0 | c orbit 1 |
| 6 | GRB 030329 | 1 | +44 | 44 | 0.07 | 0.19 | 0.24 | 0.19 | 0.3 | 1.0 | 9.8 | |
| 7 | GRB 030418 | 1 | -17 | 28 | 0.01 | -0.02 | 0.08 | 0.11 | 0.1 | -0.2 | 10.7 | |
| 8 | GRB 030429 | 1 | +46 | 29 | -0.30 | 0.21 | 0.60 | 0.40 | -0.5 | 0.5 | 8.9 | e |
| 9 | GRB 030519 | 1 | -14 | 30 | 0.55 | 0.00 | 0.67 | 0.99 | 0.8 | 0.0 | 8.4 | c |
| 10 | GRB 030528 | 1 | -14 | 37 | -1.42 | -0.40 | 1.24 | 0.55 | -1.1 | -0.7 | 8.3 | c orbits 1 & 3 |
| 11 | GRB 030723 | 1 | +5 | 28 | 0.25 | -0.42 | 0.25 | 0.21 | 1.0 | -2.0 | 9.8 | |
| 12 | GRB 030725 | 1 | -34 | 24 | 0.92 | 0.09 | 0.84 | 0.95 | 1.1 | 0.1 | 8.3 | c |
| | | 2 | | | 0.21 | -0.16 | 1.08 | 0.76 | 0.2 | -0.2 | 8.3 | c |
| 13 | GRB 030821 | 1 | -30 | 25 | -0.04 | -0.01 | 0.22 | 0.10 | -0.2 | -0.1 | 10.2 | |
| 14 | GRB 030823 | 1 | -26 | 82 | 0.82 | -0.41 | 0.39 | 0.32 | 2.1 | -1.3 | 9.3 | d |
| 15 | GRB 030824 | 1 | +19 | 41 | -0.08 | -0.14 | 0.09 | 0.09 | -0.9 | -1.6 | 10.8 | |
| 16 | GRB 030913 | 1 | +9 | 32 | -0.13 | 0.01 | 0.06 | 0.11 | -2.2 | 0.1 | 10.9 | |
| 17 | GRB 031026 | 1 | -17 | 49 | 1.10 | -0.51 | 1.25 | 0.53 | 0.9 | -1.0 | 8.3 | e, f orbit 1 |
| 18 | GRB 031203 | 2 | -11 | 22 | 4.2 | … | 0.80 | … | 5.3 | … | 8.0 | c, e |
| 19 | GRB 031220 | 1 | +18 | 70 | 0.10 | -0.33 | 0.10 | 0.11 | 1.0 | -3.0 | 10.6 | |
| 20 | GRB 040106 | 2 | +41 | 20 | -0.17 | -0.81 | 0.47 | 1.93 | -0.4 | -0.4 | 8.0 | c orbit 2, d orbit 3 |
| 21 | GRB 040323 | 2 | +1 | 21 | -0.47 | … | 0.54 | … | -0.9 | … | 8.5 | c |
| 22 | GRB 040403 | 2 | +27 | 20 | -0.53 | -0.13 | 0.27 | 0.17 | -2.0 | -0.8 | 9.8 | |
| 23 | GRB 040425 | 1 | -34 | 24 | 0.70 | -0.66 | 1.46 | 0.33 | 0.5 | -2.0 | ~9.0 | c |
| 24 | GRB 040511 | 1 | +36 | 25 | -0.33 | -0.03 | 0.22 | 0.18 | -1.5 | -0.2 | 9.9 | |
| 25 | GRB 040624 | 2 | -5 | 21 | -0.15 | -0.37 | 0.09 | 0.27 | -1.7 | -1.4 | 10.0 | |
| 26 | GRB 040701 | 1 | -34 | 24 | … | -0.07 | … | 0.56 | … | -0.1 | 8.4 | d |
| 27 | GRB 040709 | 1 | +36 | 29 | -0.13 | -0.17 | 0.46 | 0.96 | -0.3 | -0.2 | 8.5 | e |
| 28 | GRB 040730 | 2 | +18 | 24 | 0.85 | -1.24 | 1.40 | 1.22 | 0.6 | -1.0 | 7.9 | crowded galactic center |
| 29 | GRB 040810 | 1 | +3 | 24 | -0.22 | 0.05 | 0.24 | 0.14 | -0.9 | 0.4 | 10.0 | |
| | | 2 | +4 | | 0.03 | 0.34 | 0.97 | 0.20 | 0.0 | 1.7 | ~9.5 | d orbit 1, c orbit 2 |
| 30 | GRB 040812 | 2 | -29 | 20 | -0.83 | … | 0.74 | … | -1.1 | … | 8.1 | c |
| 31 | GRB 040825A | 1 | -29 | 54 | -0.61 | -0.10 | 0.41 | 0.64 | -1.5 | -0.2 | 8.9 | f |
| 32 | GRB 040825B | 1 | +3 | 74 | 0.02 | … | 0.15 | … | 0.1 | … | 9.9 | |
| 33 | GRB 040903 | 2 | +31 | 21 | -1.61 | … | 0.79 | … | -2.0 | … | 8.1 | c |
| 34 | GRB 040912 | 1 | +16 | 104 | -0.11 | 0.33 | 0.19 | 0.16 | -0.6 | 2.1 | 10.1 | |
| 35 | GRB 040916 | 1 | +6 | 74 | -1.39 | 0.42 | 3.27 | 0.67 | -0.4 | 0.6 | ~8.2 | d orbit 1, c orbit 2 |
| 36 | GRB 040924 | 1 | +34 | 42 | -2.98 | 0.02 | 0.79 | 0.10 | -3.8 | 0.2 | ~10.3 | d orbit 1 |
| 37 | GRB 041006 | 1 | +19 | 143 | -0.93 | -0.45 | 0.45 | 0.94 | -2.1 | -0.5 | 8.6 | c orbit 1, bad q's orbit 3 |
| 38 | GRB 041217 | 2 | +49 | 20 | 2.32 | -3.13 | 1.76 | 2.40 | 1.3 | -1.3 | 7.4 | d |
| 39 | GRB 041218 | 2 | +7 | 21 | 1.21 | -0.96 | 0.66 | 0.60 | 1.8 | -1.6 | 8.7 | c |
| 40 | GRB 041219A | 2 | +16 | 21 | -0.78 | 0.25 | 0.68 | 0.55 | -1.1 | 0.5 | 8.7 | e |
| 41 | GRB 041219B | 2 | +36 | 20 | -0.29 | 0.25 | 0.36 | 0.42 | -0.8 | 0.6 | 9.2 | f |
| 42 | GRB 041224 | 1 | +3 | 31 | 0.01 | -0.06 | 0.18 | 0.08 | 0.1 | -0.8 | 10.4 | |
| 43 | GRB 041226 | 2 | -22 | 22 | 0.41 | 0.04 | 0.46 | 0.14 | 0.9 | 0.3 | 9.5 | c orbit 1 |
| 44 | GRB 041228 | 3 | +8 | 25 | -1.02 | 0.55 | 1.16 | 1.19 | -0.9 | 0.5 | 8.0 | |
| 45 | GRB 050117A | 2 | +47 | 20 | -4.6 | 6.6 | 5.0 | 7.6 | -0.9 | 0.9 | 6.2 | d all 3 orbits |
| 46 | GRB 050117B | 2 | +40 | 22 | 4.5 | 0.30 | 3.05 | 7.0 | 1.5 | 0.0 | 6.4 | d all 3 orbits |
| 47 | GRB 050123 | 1 | +36 | 30 | -0.01 | 1.99 | 3.44 | 2.07 | 0.0 | 1.0 | 7.1 | d all 3 orbits |
| 48 | GRB 050124 | 2 | +43 | 22 | -0.04 | 0.23 | 0.24 | 0.20 | -0.2 | 1.2 | 9.8 | |
| 49 | GRB 050128 | 2 | +7 | 21 | 0.16 | 0.10 | 0.23 | 0.43 | 0.7 | 0.2 | 9.4 | e |
| 50 | GRB 050129 | 3 | +48 | 33 | 1.43 | -3.34 | 6.2 | 6.2 | 0.2 | -0.5 | 6.2 | |
| 51 | GRB 050219A | 1 | +37 | 26 | -0.78 | 1.92 | 1.19 | 1.78 | -0.7 | 1.1 | 7.7 | d all 3 orbits |
| 52 | GRB 050219B | 2 | +40 | 16 | -1.80 | -0.40 | 2.57 | 3.68 | -0.7 | -0.1 | 6.9 | d all 3 orbits |
| 53 | GRB 050306 | 3 | +3 | 20 | 0.25 | -1.76 | 2.42 | 2.32 | 0.1 | -0.8 | 7.2 | |
| 54 | GRB 050315 | 3 | +27 | 25 | -0.61 | -1.12 | 2.39 | 1.94 | -0.3 | -0.6 | 7.3 | |
| 55 | GRB 050318 | 3 | +9 | 26 | -2.04 | -0.02 | 0.84 | 0.88 | -2.4 | 0.0 | 8.3 | |
| 56 | GRB 050319 | 1 | -9 | 32 | 0.44 | … | 0.91 | … | 0.5 | … | 7.9 | e |
| 57 | GRB 050326 | 3 | +5 | 25 | -1.77 | 0.06 | 1.82 | 0.80 | -1.0 | 0.1 | 7.9 | |
| 58 | GRB 050401 | 1 | +13 | 24 | -0.14 | 0.08 | 0.37 | 0.29 | -0.4 | 0.3 | 9.4 | e |
| | | 2 | | | -0.18 | 0.44 | 0.20 | 0.33 | -0.9 | 1.3 | 9.6 | e |

[a] The least time in minutes between the burst and a SMEI window of observation for that location.
[b] The number of 4-second SMEI exposures during which the GRB location remains within that camera's field of view for one orbit.
[c] Modest particle contamination from SAA or auroral ovals.
[d] Substantial particle contamination from SAA or auroral ovals.
[e] Nearby bright star partly within neighboring sky bins.
[f] Nearby space debris.

We chose in figures 2 through 5 to highlight the gamma-ray burst of 2004 March 23, simply because it had the smallest $\Delta t$. As indicated by the blue patch in the middle bin of figure 5, this region of sky yields in table 2 a slightly negative value (-0.47) for the brightness of a potential optical counterpart, which lies well within the indicated range of measurement ($\sigma$ = 0.54) determined by the eight neighboring bins.

SMEI was designed to provide 0.1% differential photometry in a 1° × 1° sky bin, including photoelectron counting statistical fluctuations, sub-pixel gradients, flat-field-correction errors, and star-registration error. Here typical sky brightnesses are 100 ADUs (e.g., see figure 3), so 0.1% performance should yield a typical $\sigma$ in table 2 of $\sqrt{2}\times0.1 = 0.14$ ADU, where the $\sqrt{2}$ factor accounts the difference between two sky maps. Some $\sigma$ values are likely augmented by residual contamination from radiation-band particles, possibly some direct auroral light, and in some cases by the partial inclusion of nearby bright stars. In our example, star M Cen lies in the upper-right box of figure 5. The "comments" column of table 2 indicates which SMEI data for gamma-ray bursts are seen to suffer from particle contamination and other noise-increasing phenomena. GRB 030325 and GRB 031203 have photometric deviations greater than 3 standard deviations, but these both suffer from presumably variable bright stars within the analysis region. Finally, GRB 040924 is similarly discounted since it has significant particle contamination and also the wrong sign for a legitimate GRB when $\Delta t$ is positive.

Figure 6 presents a histogram of camera 1 and 2 $\sigma$ values from table 2, both for data having no comment and with only modest particle contamination ("c" in the table). Camera 3 is here excluded because it has a significantly higher noise level than the others. When the data are clean, SMEI's photometric performance with this type of orbit-to-orbit-difference analysis is close to the original 0.1% design specification, here corresponding to $\sqrt{2}/1000$ of sky surface brightnesses that here range from 50 to 200 ADUs. With modest contamination, $\sigma$ values become somewhat larger (shaded portion of the histogram). Typically 30% of the individual orbits do have a "noticeable" particle contamination or worse, and therefore about half of the rows in table 2 are flagged. Modest contamination residue increases $\sigma$ by two- to three-fold, but larger contamination or interference by nearby bright stars or space debris can increase $\sigma$ dramatically.

Column 12 of table 2 presents a 3-standard-deviation upper limit for $m_v$, using the preliminary estimate of one S10 = 0.4 ADU and an overall measurement standard deviation of $(\sigma_{21}+\sigma_{32})/(2\sqrt{2})$ when both differences are present, otherwise the appropriate single $\sigma$ value from the table. In five cases when one σ is much smaller than the other, the small σ was used by itself, and the upper limit marked "~". Conversion of this $m_v$ to an $m_r$ as is more usual in this context, requires an assumed spectrum shape for the alleged gamma-ray burst. When the spectrum follows frequency $\nu \propto \nu^{-1}$, then the SMEI limit for $m_r$ is about 0.25 magnitudes fainter than the $m_v$ given in table 2, whereas if $\nu \propto \nu^{-2}$, then the SMEI limit for $m_r$ is about 0.5 magnitudes fainter.

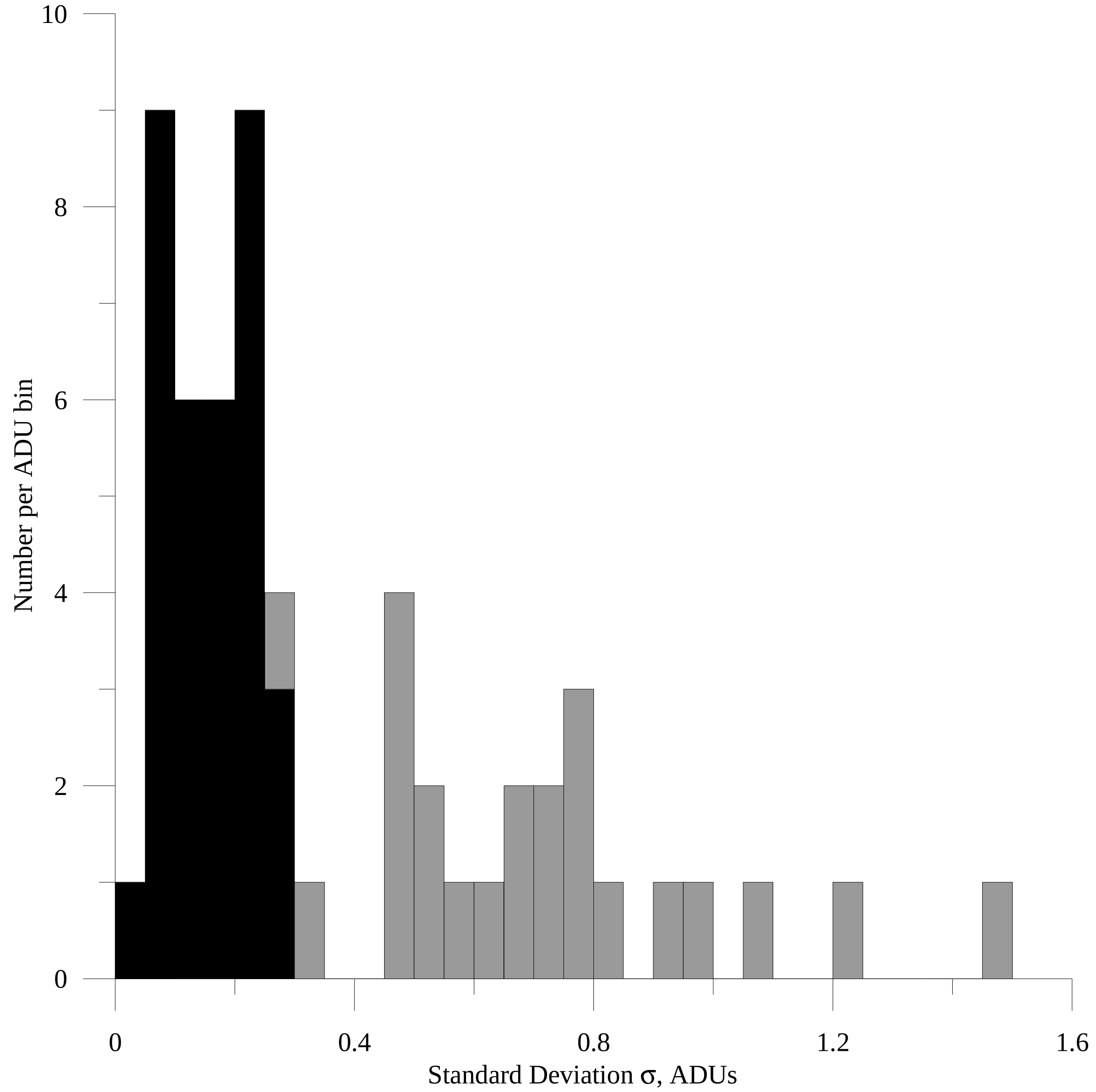

Fig. 6. –A histogram of $\sigma$ values for cameras 1 and 2 from table 2, both those free of noise-increasing background (black) and having modest residual particle contamination (gray), as indicated by comment "c".

## 5. DISCUSSION

For the study of gamma-ray afterglows, our primary result is the set of upper limits at different times Δt relative to the burst presented in Table 2. Figure 7 shows this result graphically. Assuming the intensity of the afterglow and the burst are correlated, we show in figure 8 the ratio of the optical flux upper limit normalized by the fluence as a function of Δt for those bursts where the fluence is available. In both figures the comparable values for the GRB 990123 detection (Akerlof et al. 1999) are shown. As can be seen, the optical limits are generally not particularly deep. However, the upper limits for small Δt are comparable to the GRB 990123 detection, and therefore do constrain the afterglow.

The combination of the SMEI upper limits and later afterglow detections constrain the temporal decay of the afterglow. We considered all bursts observed by SMEI within 10 minutes before or after the burst trigger. Unfortunately, no optical afterglows were detected for GRBs 030326, 040323, 040624, 040810, 040825A, 041224, 041228, 050128, 050306, and 050326. Detections or upper limits on the same timescale as, but fainter than, the SMEI observations were reported for GRBs 030723 (Smith, Akerlof, & Quimby 2003) and 050319 (Boyd et al. 2005), and therefore the SMEI observations are not interesting for these bursts. Table 3 presents the bursts for which there are both SMEI upper limits and later optical detections. Where R band detections are reported, we used a color correction of 0.5 (as discussed above) to compare the upper limits and detections. The observations constrain the power law index $\alpha$, $f_\nu \propto \Delta t^{-\alpha}$. As can be seen, none of the limits on the power law index are particularly constraining.

TABLE 3—Constraints on Afterglow Decay

| Burst | $\Delta t$[a] | $m_v$[b] | $\Delta t$[c] | Detection[d] | $\alpha$[e] | Ref.[f] |
|---|---|---|---|---|---|---|
| GRB 040916 | 6 | 8.2 | 186 | 22.3 | <4.0 | Kosugi et al. (2004) |
| GRB 041218 | 7 | 8.7 | 25 | 18.6[g] | <7.16 | Trondal (2004) |
| | | | 141 | 19.5 | <3.16 | Greco et al. (2004) |
| GRB 050318 | 9 | 8.4 | 55 | 18.4 | <4.9 | McGowan et al. (2005) |

[a] Time between burst trigger and SMEI observation, in minutes.
[b] SMEI $3\sigma$ upper limit.
[c] Time in minutes between burst trigger and optical detection
[d] Optical detection, in $m_r$, unless otherwise specified.
[e] Power law index $\alpha$, $f_\nu \propto \Delta t^{-\alpha}$, of the afterglow's temporal decay.
[f] Reference for optical detection.
[g] Optical detection in V band.

The black parts of the histogram in figure 6 show -- in the absence of interference by Earth radiation- band particles, space debris, or variation due to nearby stars -- that SMEI imagery is indeed differentially photometric at the design specification of 0.1%, within a 1° × 1° sky bin. This performance was noted in figure 8 of Jackson et al. (2004), but here extends to the wide variety of sky locations and times of these gamma-ray bursts.

While the SMEI sensitivity cannot compare with the sensitivity of the UVOT on *Swift* or the ground-based robotic telescopes such as ROTSE-III, future SMEI observations are still relevant when these more sensitive telescopes are unable to observe a burst, whether because of observational constraints or because of bad weather. SMEI observations, whether detections or upper limits, are relevant given the optical detections of GRB 990123 and, although fainter, of GRB 041219A and given the complex X-ray afterglows observed by *Swift*'s XRT. Therefore, from time to time we will release detections and/or upper limits for the burst locations viewed by SMEI in the hour before or after the burst.

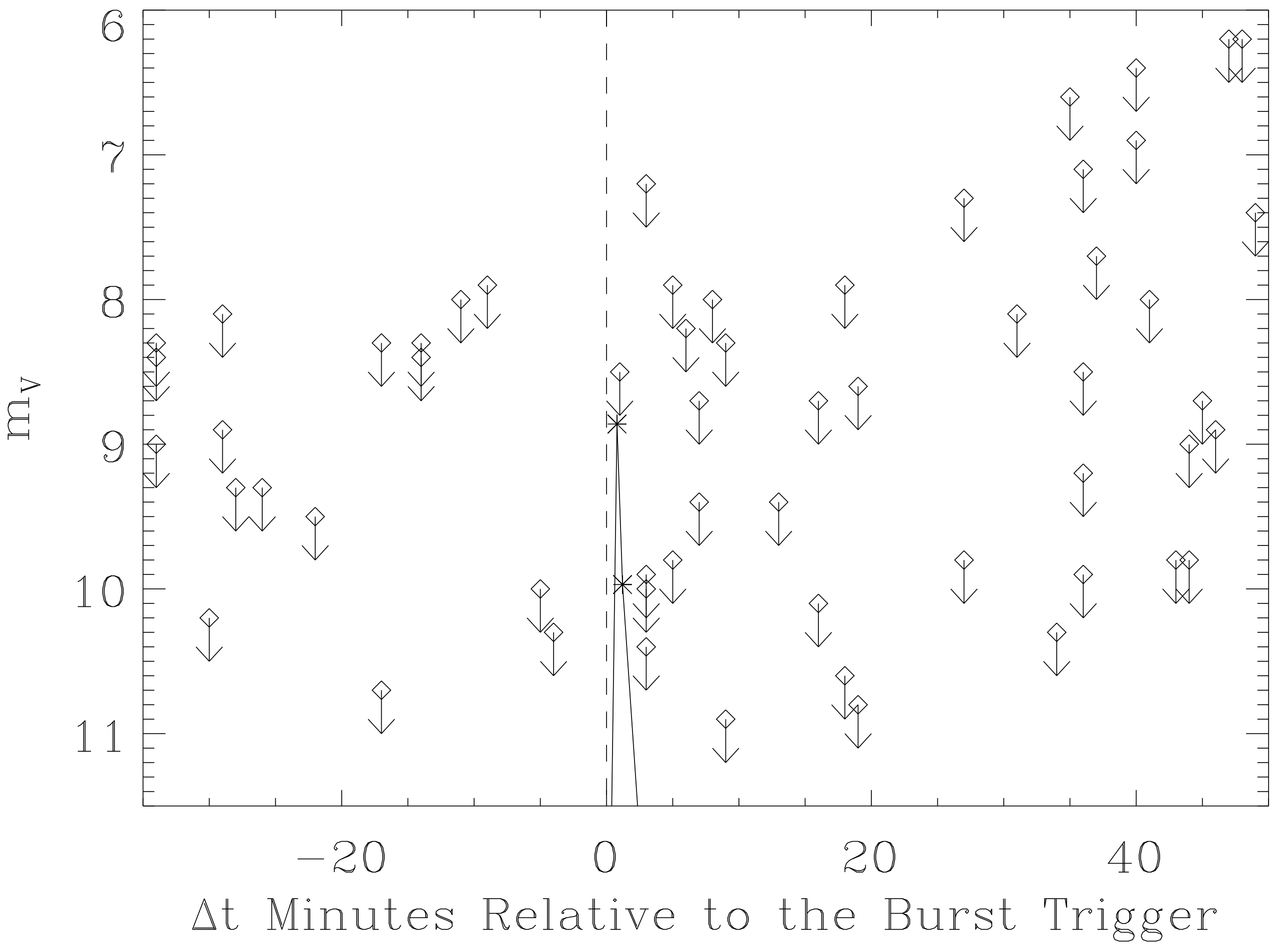

Fig. 7. –SMEI 3σ upper limits for $m_v$ vs. Δt (time relative to the burst), from table 2. The points with ‘*’ connected by solid lines are the GRB 990123 observations (Akerlof et al. 1999).

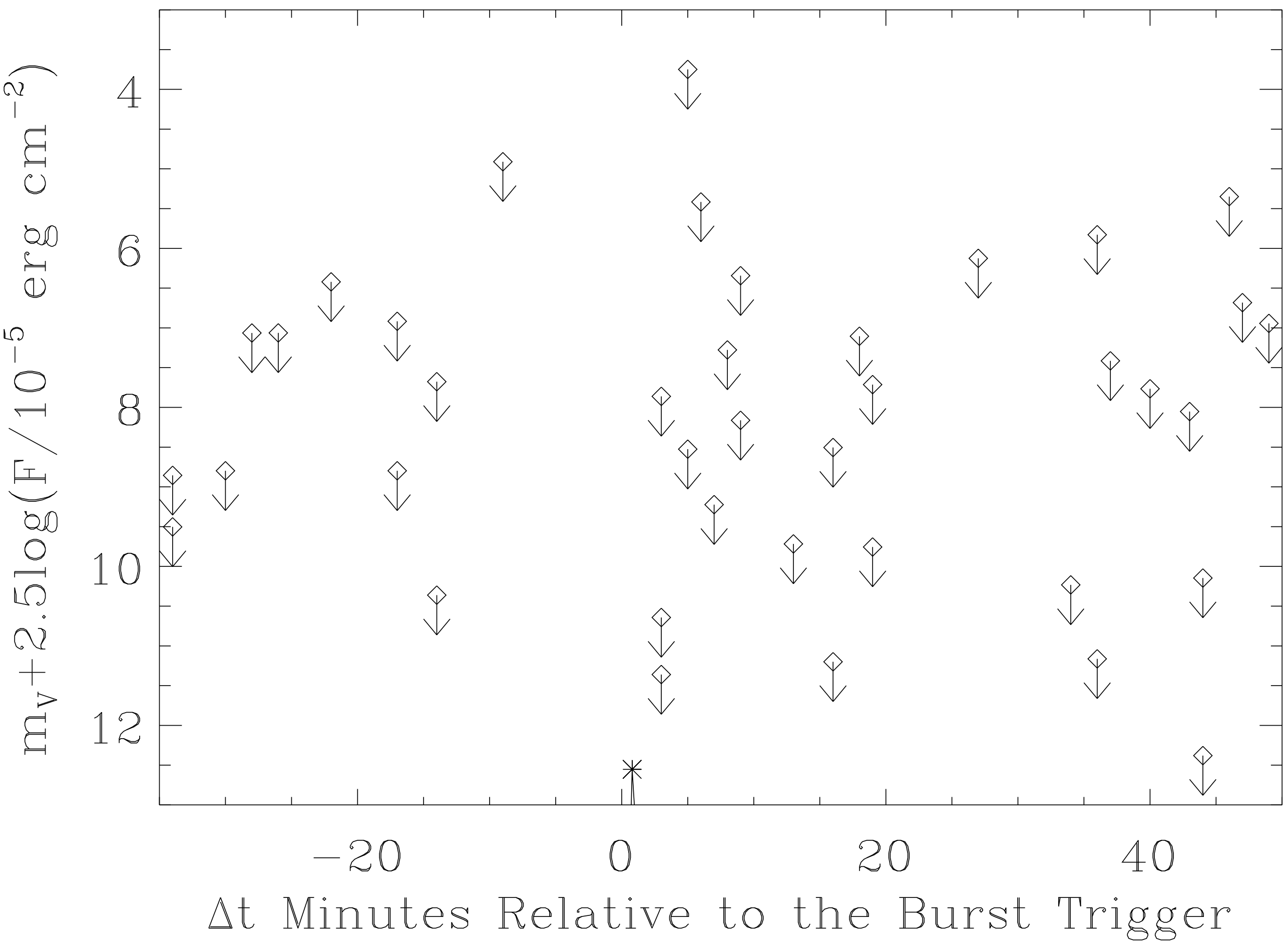

Fig. 8. –SMEI 3σ upper limits for $m_V$ normalized by the fluence F in units of $10^{-5}$ erg $cm^{-2}$ (i.e., $m_V$+2.5log[F/ $10^{-5}$ erg $cm^{-2}$]) vs. Δt. The point with '*' near Δt=0 is the GRB 990123 observation (Akerlof et al. 1999).

## 6. ACKNOWLEDGEMENTS

SMEI was designed and constructed by a team of scientists and engineers from the U.S. Air Force Research Laboratory, the University of California at San Diego, Boston College, Boston University, and the University of Birmingham in the U.K. Financial support was provided by the Air Force, the University of Birmingham, and NASA. The work at UCSD was supported in part by AFRL contract AF19628-00-C-0029, AFOSR contract AF49620-01-0054, and NASA grant NAG5-134543.